\documentclass[aps,showpacs,preprintnumbers,amsmath,amssymb]{revtex4-2}
\usepackage{amsmath,amssymb,graphics,epsfig,subfigure}
\usepackage{color}
\usepackage{multirow}
\usepackage{graphicx}
\usepackage{epstopdf}
\usepackage{dcolumn}
\usepackage{amsmath, mathrsfs}
\usepackage{amsmath}
\usepackage{amssymb}
\usepackage{bm}
\usepackage{color}
\usepackage{multirow}
\usepackage{float}
\usepackage[dvipsnames]{xcolor}
\usepackage{hyperref}

\begin{document}
\newcommand {\nn} {\nonumber}
\renewcommand{\baselinestretch}{1.3}

 \baselineskip=0.8cm
\title{Precessions of spherical orbits in the rotating Melvin black hole spacetime and its constraints from the jet of M87*}

\author{Chengjia Chen$^{1}$, Qiyuan Pan$^{1,2}$\footnote{panqiyuan@hunnu.edu.cn}, and Jiliang Jing$^{1}$\footnote{jljing@hunnu.edu.cn}}

\affiliation{$^1$Department of Physics, Institute of Interdisciplinary Studies, Key Laboratory of Low Dimensional Quantum Structures
    and Quantum Control of Ministry of Education, Synergetic Innovation Center for Quantum Effects and Applications, Hunan
    Normal University,  Changsha, Hunan 410081, People's Republic of China
    \\
    $ ^2$Center for Gravitation and Cosmology, College of Physical Science and Technology, Yangzhou University, Yangzhou 225009, People's Republic of China}

\begin{abstract}
 \baselineskip=0.6cm

We investigate precessions of spherical orbits of timelike particles in the background of a rotating black hole immersed in the Melvin magnetic field, and probe effects of the magnetic field on the precession period. Our results show that effects of the magnetic field on the particles' motions gradually decrease with the titled angle and finally vanish as the titled angle tends to $\pi/2$. With the increase of the  magnetic field parameter, we find that the precession becomes rapidly.  Modelling the spherical orbit to the warp radius in the accretion disk and with the observed precession period from the jet of  M87*, we analyze the allowed regions of the black hole parameters and the warp radius in the accretion disk. Especially, we find a novel degenerated phenomenon of precession periods arising from magnetic field for two different spherical orbits, which does not appear in the usual Kerr black hole case. Moreover, we also discuss the possibility of observing effects of the magnetic field on the precession periods of jets for astrophysical black holes. Our study could help to further understand the rotating Melvin black hole and the relationship between magnetic fields and precessions of jets.

\end{abstract}


\maketitle

\newpage
\section{Introduction}
\label{secIntroduction}

 The first direct image of the supermassive black hole in the M87 Galaxy \cite{Akiyama} was released by the Event Horizon Telescope (EHT) Collaboration, which is a milestone event in physics because it provides direct visual evidence of black hole in our universe and confirms that black hole is no longer just a theoretical model. Accretion disks as well as jets near black holes, are actual light sources illuminating black holes due to their strong electromagnetic
emissions. It is undoubted that the matter configuration including accretion disks and jets are indelibly imprinted on black
hole images. Conversely, analyzing the luminosity distribution and electromagnetic signals stored in the images can extract the information about accretion disks and jets near black holes. Therefore, a lot of efforts have been devoted to the study of black hole images from various aspects \cite{Akiyamab,Kocherlakota,kerr1, kerr2,bhs0,bhs1,bhs2,bhs3,bhsp1,sw,swo,astro,chaotic,Banerjee}. Recently, the high-resolution images of M87* at a wavelength of 3.5 mm \cite{Lu} revealed the a triple-ridge jet structure near the black hole besides a ring-like structure with a diameter of approximately 8.4 Schwarzschild radii, which further confirms the connection between the jet and the accretion flow.
A remarkable finding \cite{Yuzhu} obtained by analysing of the radio galaxy M87 over $22$ years reveals that there is a periodic variation in the position angle of the jet with a period of approximately 11 years and the corresponding half-opening angle of the precession cone is about to $1.25\pm0.18$ degrees. This observation can be interpreted by the Lense-Thirring precession due to the misalignment of the orbital angular momentum and the black hole spin. Therefore, the supermassive black hole  M87* galaxy must be a spinning black hole with a tilted accretion disk. The theoretical studies of tilted accretion disks have been extensively explored \cite{Petterson, Ostriker,Lodato,Zahrani}. The quasi-periodic brightness oscillations related to tilted accretion disks are also analyzed in \cite{Fragile}. The spherical orbits whose orbital planes deviated from equatorial plane have been analysed in the Kerr black hole \cite{constraining}. The precessions of the black hole jets are modelled by those of spherical orbital planes based on three assumptions, and then the black hole's parameters was constrained by using the observed precession period of the M87 jet \cite{constraining}, which also provides an alternative way to examine theory of gravity in the strong field regime.

In real astrophysical situations, there may exist an external magnetic field around supermassive black holes  at the center of galaxies. For example, the magnetic field strength around the M87* is estimated as $B\approx 1-30 Gs$ \cite{EHT1,EHT2}, and  the magnetic field strength around the Seyfert galaxy MCG-6-30-15 reaches up to $B\approx 10^{10}-10^{11} Gs$ \cite{Zakharov}. Therefore, it is much interesting to research observable effects of black holes arising from magnetic fields. The solutions of a Schwarzschild black hole and rotating black hole immersed in uniform  magnetic fields were respectively obtained by Ernst et al \cite{lv00,lv01}, whose magnetic field is cylindrically symmetrical and is aligned along the symmetry axis of the black hole. The presence of the magnetic field makes the spacetime asymptotically non-flat at the spatial infinite. Although their positions of the event horizon are the same as in the usual black holes without magnetic field, the polar circumference for the event horizon increases with the magnetic field, while the equatorial circumference decreases. Moreover, chaotic motions of test particle occur in the Melvin black hole spacetimes because the geodesic equations are not variable-separated due to the existence of magnetic field \cite{chaom1,chaom2}. The black hole shadows \cite{chaomsha1,chaomsha2} are found to own certain fractal structures caused by chaotic lensing. Moreover, the shadow could be a panoramic (equatorial) shadow for the magnetic field parameter exceeding a certain threshold \cite{chaomsha1,chaomsha2}.  The main purpose of this paper
is to probe precessions of spherical orbital planes in the rotating Melvin black hole spacetime and to ask what new effects arising from the magnetic field together with the titled angle of the orbital plane.
As in \cite{constraining}, we will model the spherical orbital radius to the warp radius \cite{Petterson,Ostriker} and discuss possible constraints on black hole parameters by using the observed precession of the M87* jet nozzle. The main reason of selecting the warp radius as characteristic radius  \cite{Petterson,Ostriker} the accretion disk rather than the innermost stable circular orbit (ISCO) is that the disk at the warp radius is pulled back into the equatorial plane, and particles exceed the innermost stable circular ISCO must undergo a rapid fall into the black hole.

The paper is organized as follows: In Sec.\ref{dddd}, we  present an analysis of spherical orbits of timelike particles around a rotating black hole immersed in Melvin magnetic field, and probe effects of the magnetic field on these orbits. 
In Sec.\ref{ptco}, we analyze the precessions of spherical orbital planes in the rotating Melvin black hole spacetime and discuss the possibility of constraint on black hole parameters by the observed period of the precession. Finally, we end the paper with a summary.

\section{Spherical orbits around a rotating black hole immersed in Melvin magnetic field }
\label{dddd}

  Let us now to review briefly on the rotating black hole immersed in a strong uniform magnetic field, which is an axially symmetric and non-asymptotically flat solution of the Einstein-Maxwell equations with the magnetic field aligned along its symmetry axis. With the standard Boyer-Lindquist coordinates, its metric has a form  \cite{lv01}
\begin{eqnarray}
\label{mkdg}
ds^{2}=|\Lambda|^{2}\Sigma [-\frac{\Delta}{\mathcal{A}}dt^{2}+\frac{dr^{2}}{\Delta}+d\theta^{2}]+\frac{\mathcal{A}}{\Sigma|\Lambda|^{2}}\sin^{2}\theta(d\phi-\omega dt)^{2},
\end{eqnarray}
where
\begin{eqnarray}
\label{mkdg1}
\Sigma &=& r^{2}+a^{2} \cos^{2}\theta,\;\;\;\;\;\;\;\;\;\;\; \mathcal{A}=(r^{2}+a^{2})^{2}-\Delta a^{2} \sin^{2}\theta, \nonumber\\ \;\;\;\;\;\;\
\Delta &=& r^{2}-2 Mr+a^{2}, \;\;\;\;\;\;\;\;\; \Lambda=1+\frac{B^{2}\mathcal{A}}{4\Sigma}\sin^{2}\theta-\frac{i}{2}B^{2}Ma \cos\theta (3-\cos^{2}\theta+\frac{a^{2}}{\Sigma}\sin^{4}\theta),
\end{eqnarray}
and
\begin{eqnarray}
\label{mkdg2}
\omega&=&\frac{a}{r^{2}+a^{2}}\Bigg\{(1-B^{4}M^{2}a^{2})-\Delta\Bigg[\frac{\Sigma}{\mathcal{A}}+\frac{B^{4}}{16}\Bigg(-8Mr\cos^{2}\theta(3-\cos^{2}\theta)-6Mr\sin^{4}\theta\nonumber\\
&+&\frac{2Ma^{2}\sin^{6}\theta}{\mathcal{A}}[r(r^{2}+a^{2})+2Ma^{2}]+\frac{4M^{2}a^{2}\cos^{2}\theta}{\mathcal{A}}[(r^{2}+a^{2})(3-\cos^{2}\theta)^{2}-4a^{2}\sin^{2}\theta]\Bigg)\Bigg]\Bigg\}.
\end{eqnarray}
Here  $M$ and $a$ respectively correspond to the mass and spin parameters of the black hole, and the parameter $B $ describes the magnetic field strength. The metric (\ref{mkdg}) reduces to the Kerr case as the parameter $B =0$, and to the Melvin-Schwarzschild one as $a=0$.
From Eq.(\ref{mkdg}), one can fin that the black hole horizons  are located at
\begin{eqnarray}
\label{horizon}
r_{\pm}=M\pm\sqrt{M^{2}-a^{2}},
\end{eqnarray}
which are the same as those of the Kerr black hole. Near the horizon, the spacetime geometry is found to be contributed by the strong gravity mainly arising from the black hole mass. However, the presence of the parameter $B$ yields that the spacetime is non-asymptotically flat as in the Melvin universe,
which is different from that in the Kerr black hole spacetime.

We are now in position to study the motion of massive particle in the Melvin-Kerr black hole spacetime (\ref{mkdg}). The equation of motion for a massive particle moving along timelike geodesics in curved spacetime can be derived from the Lagrangian equation, where the Lagrangian can be expressed as
\begin{eqnarray}
\mathcal{L}=\frac{1}{2}g_{\mu\nu}\dot{x^{\mu}}\dot{x^{\nu}}.
\label{Lagrangian}
\end{eqnarray}
Obviously, the metric functions in the Melvin-Kerr black hole spacetime(\ref{mkdg}) are independent of the coordinates $t$ and $\psi$. Therefore, the massive particle's energy $E$ and its $z$-component of the angular momentum $L$ are two conserved quantities, and their expressions can be respectively expressed by
\begin{eqnarray}
E=g_{tt}\dot{t}+g_{t\varphi}\dot{\varphi},\quad \quad\quad L=g_{\varphi\varphi}\dot{\varphi}+g_{t\varphi}\dot{t}.
\label{energy}
\end{eqnarray}

With these two conserved quantities, one can obtain the timelike geodesics equations
\begin{eqnarray}
 \dot{t}&=&\frac{g_{\varphi\varphi}E+g_{t\varphi} L}{g_{t\varphi}^2-g_{tt}g_{\varphi\varphi}},\label{radia}\\
\dot{\varphi}&=&-\frac{g_{t\varphi}E+g_{tt} L}{g_{t\varphi}^2-g_{tt}g_{\varphi\varphi}},\label{tth}\\
\frac{d}{d\lambda}(g_{rr}\dot{r})&=&\frac{1}{2}\bigg[(\partial_{r}g_{tt})\dot{t}^2+2(\partial_{r}g_{t\varphi})\dot{t}\dot{\varphi}
+(\partial_{r}g_{\varphi\varphi})\dot{\varphi}^2+
2(\partial_{r}g_{rr})\dot{r}^2+(\partial_{r}g_{\theta\theta})\dot{\theta}^2\bigg],\label{cedx0r}\\
\frac{d}{d\lambda}(g_{\theta\theta}\dot{\theta})&=&\frac{1}{2}\bigg[(\partial_{\theta}g_{tt})\dot{t}^2+2(\partial_{\theta}g_{t\varphi})\dot{t}\dot{\varphi}
+(\partial_{\theta}g_{\varphi\varphi})\dot{\varphi}^2+
(\partial_{\theta}g_{rr})\dot{r}^2+2(\partial_{\theta}g_{\theta\theta})\dot{\theta}^2\bigg],\label{cedx0th}
\end{eqnarray}
with the constrain condition $u^{\mu}u_{\mu}=-1$, i.e.,
\begin{eqnarray}
g_{rr}\dot{r}^2+g_{\theta\theta}\dot{\theta}^2&=&V_{eff}(r,\theta; E,L),\label{u3}
\end{eqnarray}
where  the effective potential $V_{eff}(r,\theta; E,L)$ has a form
\begin{eqnarray}
V_{eff}(r,\theta; E,L)=\frac{E^2g_{\varphi\varphi}+2ELg_{t\varphi}+L^2g_{tt}
}{g^2_{t\varphi}-g_{tt}g_{\varphi\varphi}}-1.
\end{eqnarray}
For the particle moving along the spherical orbits deviating from the equatorial plane, its polar angle  $\theta$ is found to be changed and  the value of $\theta$ is confined within the range $[\frac{\pi}{2}-\xi,\frac{\pi}{2}+\xi]$, where  $\xi$  is the so-called tilt angle described
the maximum half-opening angle of the motion along the $\theta$ direction. At the turning point $\theta=\frac{\pi}{2}-\xi$ or $\theta=\frac{\pi}{2}+\xi$, one has $\dot{r}=\dot{\theta}=\ddot{r}=0$, which means that in this case Eq.(\ref{cedx0r}) can be simplified as \cite{motta1,motta2,motta3,motta4}
\begin{eqnarray}\label{tdot0}
(\partial_{r}g_{tt})\dot{t}^2+2(\partial_{r}g_{t\varphi})\dot{t}\dot{\varphi}+(\partial_{r}g_{\varphi\varphi})\dot{\varphi}^2=0.
\end{eqnarray}
Thus, the orbital angular velocity $\Omega_{\varphi}$ of particle at the turning point has a form 
\begin{eqnarray}
\Omega_{\varphi}=\frac{d\varphi}{dt}=\frac{-g_{t\varphi,r}\pm\sqrt{(g_{t\varphi,r})^2
+g_{tt,r}g_{\varphi\varphi,r}}}{g_{\varphi\varphi,r}},\label{jsd0}
\end{eqnarray}
where the sign is $+ (-)$ for corotating (counterrotating)
orbits. From the condition $g_{\mu\nu}\dot{x}^{\mu}\dot{x}^{\nu}=-1$, at the turning point with $\dot{r}=\dot{\theta}=0$, one has
 another  independent relationship between $\dot{t}$ and $\dot{\varphi}$
\begin{eqnarray}\label{tdot1}
g_{tt}\dot{t}^2+2g_{t\varphi}\dot{t}\dot{\varphi}+g_{\varphi\varphi}\dot{\varphi}^2=-1.
\end{eqnarray}
Making use of two equations (\ref{tdot0}) and (\ref{tdot1}), one can obtain 
\begin{eqnarray}\label{t0}
\dot{t}=\frac{1}{\sqrt{-g_{tt}-2g_{t\varphi}\Omega_{\varphi}-g_{\varphi\varphi}\Omega_{\varphi}^2}}.
\end{eqnarray}
Inserting Eq.(\ref{radia}) and (\ref{tth}) with the condition $\dot{r}=0$, one can get the energy $E$ and the angular momentum $L$ of massive particles along  the spherical orbits 
\begin{eqnarray}
&&E=-\frac{g_{tt}+g_{t\varphi}\Omega_{\varphi}}{\sqrt{-g_{tt}-2g_{t\varphi}\Omega_{\varphi}
-g_{\varphi\varphi}\Omega^2_{\varphi}}},\nonumber\\
&&L=\frac{g_{t\varphi}+g_{\varphi\varphi}\Omega_{\varphi}}{\sqrt{-g_{tt}
-2g_{t\varphi}\Omega_{\varphi}-g_{\varphi\varphi}\Omega^2_{\varphi}}}.\label{jsd}
\end{eqnarray}
\begin{figure}[ht]
\includegraphics[width=6cm]{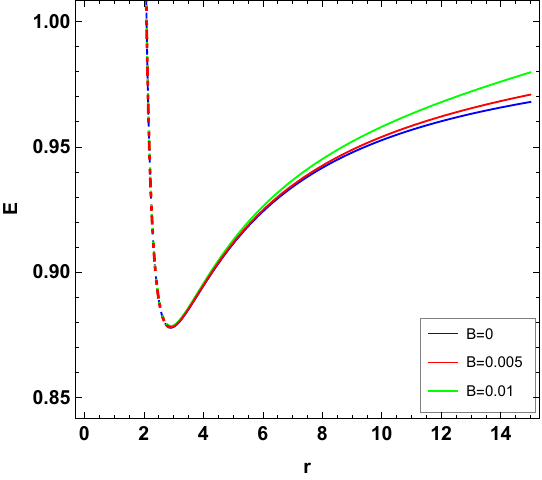}\quad \includegraphics[width=5.9cm]{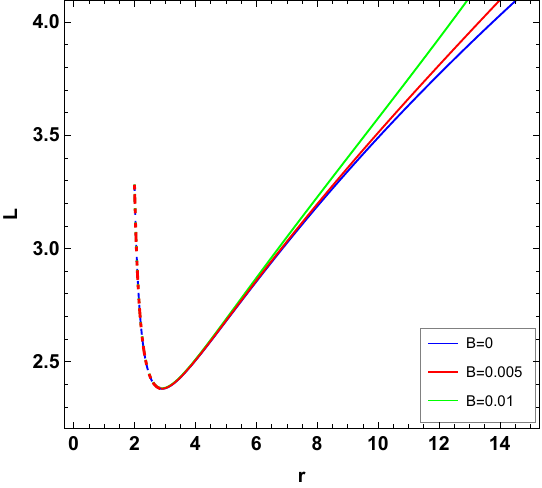}\\
\includegraphics[width=6cm]{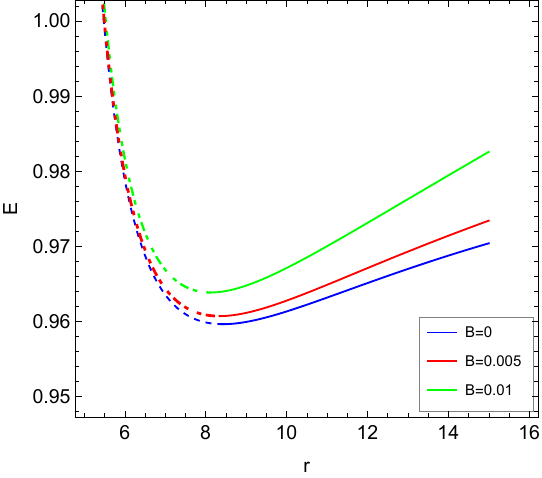}\quad \includegraphics[width=5.9cm]{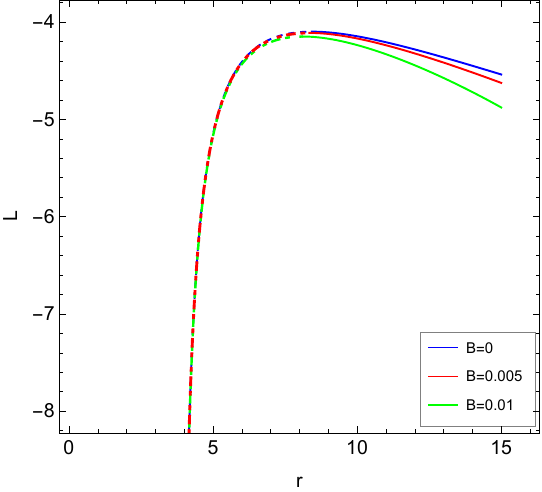}
\caption{Varieties of energy $E$ and angular momentum $L$ with the radius $r$ of a circular orbit and the magnetic
field parameter $B$ for fixed $\xi=\pi/144$. Panels in the upper row are for the prograde orbits and in lower row are for the retrograde orbits. Here we set $a=0.8$ and $M=1$.}
\label{sx1}
\end{figure}
Fig. \ref{sx1} show the variations of  energy $E$ and angular momentum $L$  with the radii $r$ of spherical orbits and the magnetic field parameter $B$. In each panel, dashed and solid curves correspond to unstable and stable orbits, respectively. The intersection point between dashed and solid curves corresponds to the innermost stable circular orbit. As in the Kerr case, the unstable circular orbits also possess a minimum radius, which yields that the energy and angular momentum are divergent. For the stable circular orbits outsides the ISCO,  their energy of always are less than 1. The presence of the magnetic field strength $B$ yields that both the orbital energy $E$ and the angular momentum's absolute values $|L|$ increase in the prograde or retrograde cases.
\begin{figure}[ht]
\includegraphics[width=5.8cm]{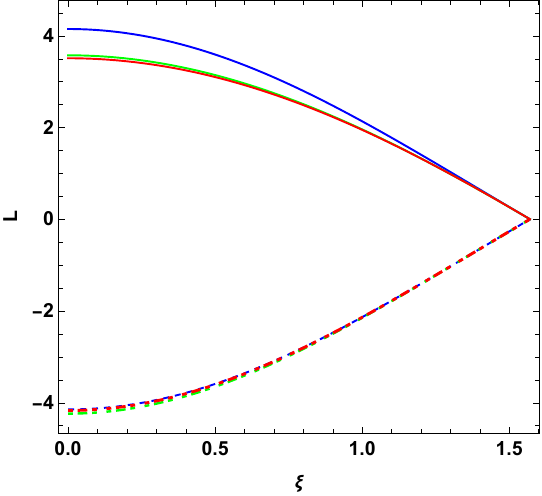}\quad \includegraphics[width=6cm]{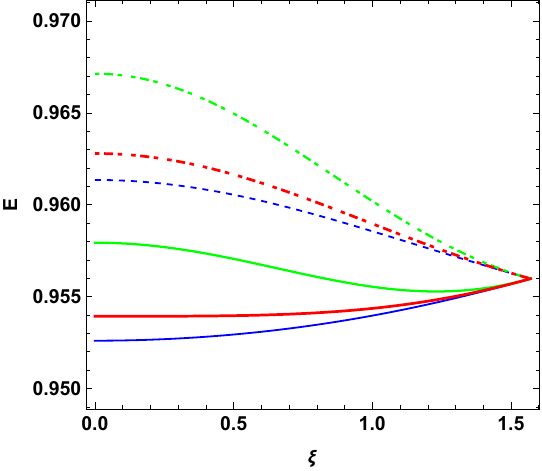}
\caption{Varieties of energy $E$ and angular momentum $L$ with the tilt angle $\xi$ for different magnetic
field parameter $B$. The solid curves and dashed curves are for the prograde and retrograde circular
orbits. The curves with blue, green and red respectively correspond to the cases with $B=0$, $0.005$ and $0.01$. Here we set $a=0.8$, $r=10$ and $M=1$.}
\label{sx2}
\end{figure}
Moreover, we also find that the effects of magnetic field on $|L|$ of particles moving along the prograde orbits are larger than those along the retrograde orbits. From Fig. \ref{sx2}, we find the magnetic field affects the dependence of the angular momentum and energy on the tilt angle. It is shown that effects of magnetic field decrease with the tilt angle $\xi$ and finally vanish as $\xi=\frac{\pi}{2}$.

\begin{figure}[ht]
\includegraphics[width=6cm]{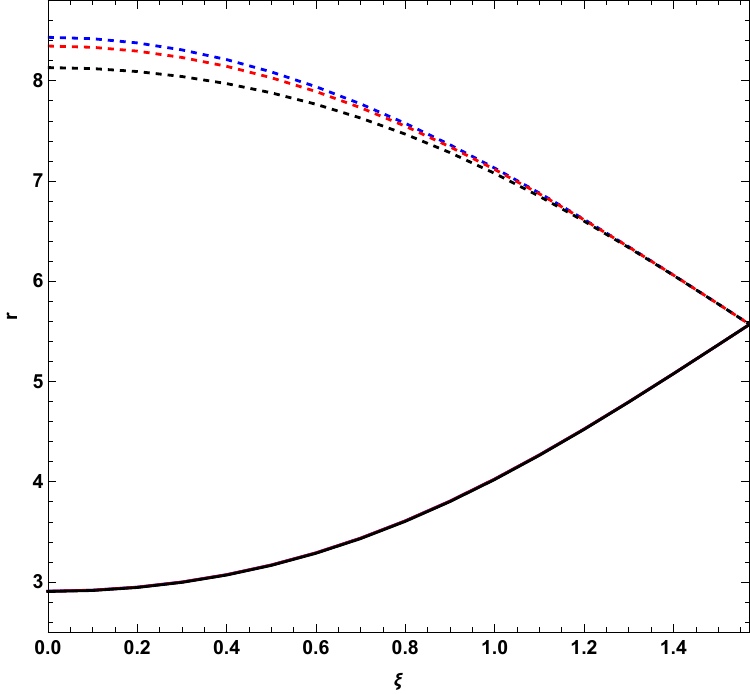}
\caption{ Changes of the ISCO radius with $\xi$ for fixed $a=0.8$. The blue, red and black curves respectively correspond to the case with $B=0$, $0.01$,$0.02$. Dashed and solid curves are for the retrograde and prograde ISCOs. Here we set $M=1$.}
\label{sx3}
\end{figure}

As in \cite{constraining}, the ISCO can be equivalently determined by
\begin{eqnarray}
 \left(\frac{dL}{dr}\right)_{a,\xi}=0, \quad \quad\quad \left(\frac{dE}{dr}\right)_{a,\xi}=0.
\end{eqnarray}
The dependence of the radius $r$ of $ISCO$ on the tilt angle $\xi$ for different magnetic field parameter $B$ is shown in Fig. \ref{sx3}, which reveals that both the retrograde and prograde innermost stable circular orbital radii decrease with the increasing magnetic field parameter $B$.
Similarly, effects of magnetic field on the ISCO radius decrease with the tilt angle $\xi$ and finally vanish as $\xi=\frac{\pi}{2}$. The effects of magnetic field on the ISCO radius of particles moving along the prograde orbits are shown to be less than those along the retrograde orbits in Fig. \ref{sx3}.

\section{ Precession of spherical orbits and constrains from M87*}
\label{ptco}

\begin{figure}[ht]
\centering
\includegraphics[width=4cm]{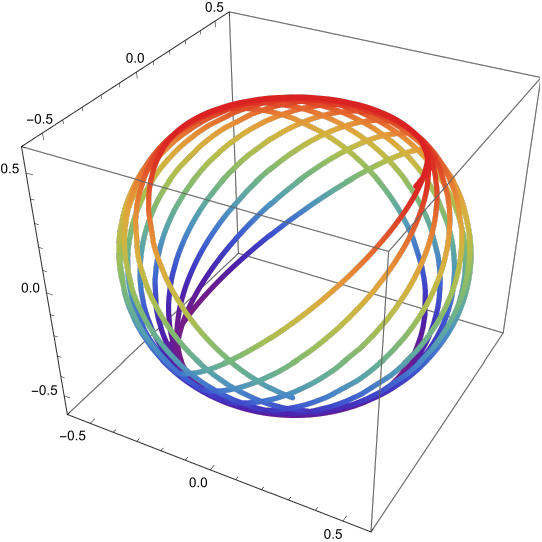}\;\;\;\includegraphics[width=4cm]{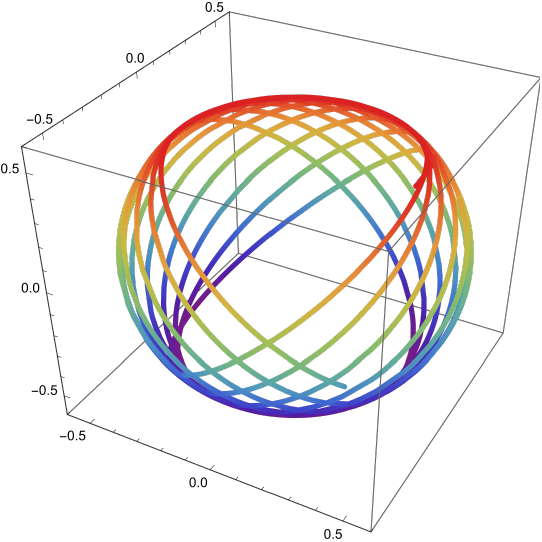}\;\;\;\includegraphics[width=4cm]{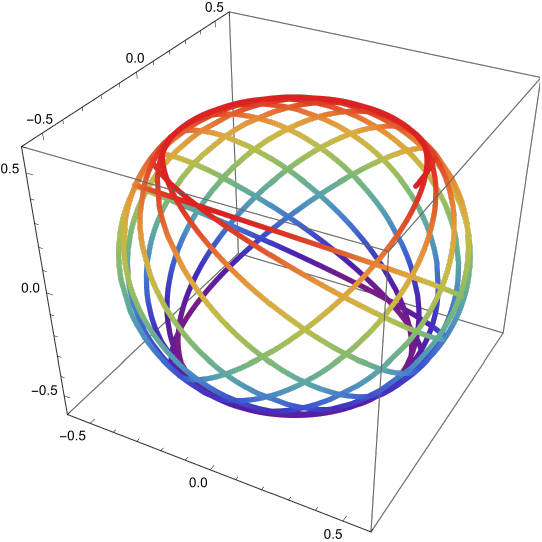}
\caption{The trajectory of a massive particle moving along a tilted circular orbit with $r=10$ and $\xi=\pi/4$ for different $B$. Panels from left to right respectively denote the cases $B=0$, $0.005$ and $0.01$. Here we set $M=1$ and $a=0.8$.}
\label{trajectory}
\end{figure}

For the spherical orbits, there exist the Lense-Thirring precession due to existence of an angle between the directions of the orbital angular momentum and the black hole spin, which yields that the orbit plane deviates from its initial plane. In this section, we will focus on studying the effect of the magnetic field on the precession of spherical orbits. Fig. \ref{trajectory} shows trajectory diagram of a particle with mass moving along a spherical orbit with an inclined angle $\xi=\pi/4$ and the orbital radius $r=10$. And the black hole spin is set to $a=0.8$.
The particle initially starts from the top turning point with $\theta=\pi/2-\xi$ and gradually moves towards the bottom turning point with $\theta=\pi/2+\xi$, and then makes a U-turn and proceeds in the northeast direction. However, the  direction of the line connecting two turning points is changing with time, which leading to a precession of the orbital plane. Fig. \ref{trajectory} shows that the precession becomes rapidly with the increase of the  magnetic field parameter $B$. The precession angular velocity $\omega_{t}$ of the tilted circular orbit can be calculated with \cite{constraining}
 \begin{eqnarray}
 \omega_{t}=\frac{\Delta\varphi-2\pi}{T_{\theta}},
 \label{velocity}
\end{eqnarray}
 where $T_{\theta}$ denotes the period of the $\theta$ motion, and the $\Delta\varphi$  describes the increment in angular $\varphi$ over one period of the $\theta$ motion. Here, the precession angular velocity $\omega_{t}$ is for the distant observer rather the local observer. In Tab. \ref{tab}, we present
 precession angular velocity $\omega_{t}$ of particles moving along spherical orbits with different magnetic parameter $B$.
 \begin{table}[ht]
\setlength{\tabcolsep}{2.5mm}{\begin{tabular}{|c|c|c|}\hline\hline
  \multirow{2}{*}{\;}    & \multicolumn{2}{c|}{$\omega_t$}       \\ \cline{2-3}
      &$\xi=\pi/4$&$\xi=4\pi/9$ \\\hline
$B=0$      & 0.00137568 &  0.00153188  \\\hline
 $B=0.005$     & 0.00153417 & 0.00157015    \\\hline
$B=0.01$      & 0.00200991 & 0.00168542       \\\hline\hline
\end{tabular}
\caption{Precession angular velocity $\omega_{t}$ of particles moving along spherical orbits with different magnetic parameter $B$. We set $M=1$, $a=0.8$ and $r=10$.}\label{tab}}
\end{table}
Notably, the presence of the magnetic field parameter $B$ enhances the precession angular velocity $\omega_{t}$ and then shortens  precession period $T$, which is consistent with that obtained in Fig. \ref{trajectory}.

Recently,  a notable precession with $11.24 \pm0.47 $ years is reported by Cui et al. \cite{Yuzhu} for the jet axis of $M87^*$, which means that the accretion disk around the black hole $M87^*$ is a tilted disk deviated from the equatorial plane. It is believed that the characteristic radius of the whole titled disk is represented by the warp radius rather than ISSO \cite{Petterson,Ostriker}. The main reason is that particles approaching the ISSO rapidly plunge into the central black hole and at the warp radius the tilt angle of the accretion disk  vanishes. Thus, it is widely believed that the jet originates from the warp radius within the range of $ (6M, 20M)$ \cite{constraining}.
Modelling the spherical orbit to the warp radius in the accretion disk, the constraints on Kerr black hole's parameters was studied by using the observed precession period \cite{constraining},
which provides an explicit relationship between the black hole spin and the warp radius in the allowed region.
\begin{figure}
\includegraphics[width=15cm]{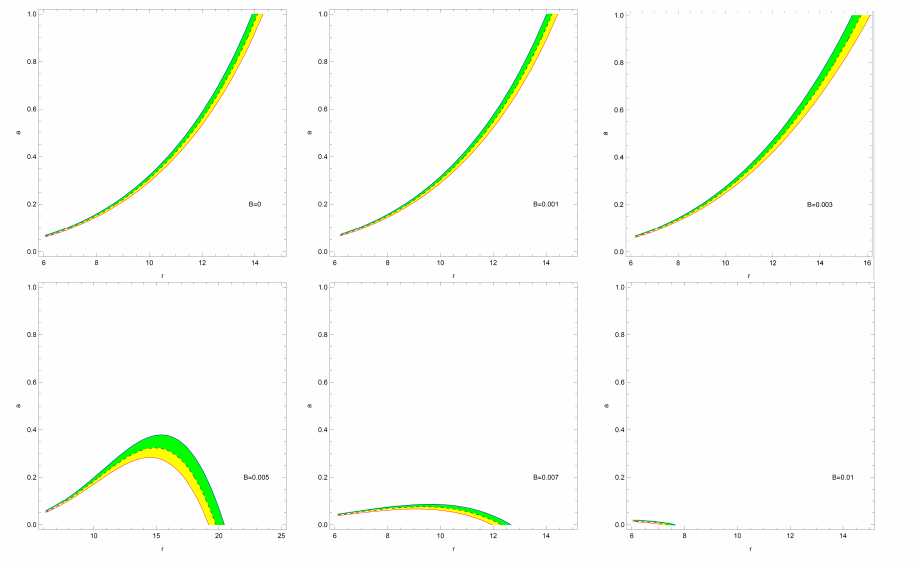}
\caption{Constrains of the black hole spin and warp radius of the accretion disk with the precessing period of the jet nozzle of M87* for different magnetic field parameter $B$ in the prograde cases. The tilt angle $\xi=\pi/144=1.25^{\circ}$. The solid curve corresponds the period $T=11.24$ years. The blue and red dashed curves are for $T=10.77$ and $11.71$ years. }
\label{sx5}
\end{figure}
Along this way, we will discuss the corresponding constraints on the rotating black hole immersed in Melvin magnetic field. The period of the precession for the tilted circular orbit around the black hole M87* can be expressed as \cite{constraining}
\begin{eqnarray}
 T=\frac{2\pi}{\omega_{t}}\frac{GM_{\odot}}{c^{3}}\left(\frac{M}{M_{\odot}}\right)\approx 6.394\times 10^{-3}\times\frac{1}{\omega_{t}}\;\;(yr),
   \label{ period}
\end{eqnarray}
where the unit is restored and $M=6.5\times10^9M_{\odot}$. With the fixed tilt angle $\xi=\pi/144=1.25^{\circ}$ and the observed precession period from M87*, we
present the allowed regions of the black hole spin $a$ and the warp radius $r$ in the accretion disk for different magnetic field parameter $B$.
From Fig. \ref{sx5}, for the prograde cases with $a>0$, as the parameter $B$ is smaller, one can find that
for each fixed period the value of the black hole spin increases with the warp radius which is similar to that in the Kerr case. We also find that the warp radius increases with the parameter $B$ for fixed $a$ in this small $B$ case.
With the gradual increasing of the parameter $B$, the value of the black hole spin first increases and then decreases with the warp radius. With the further increase of $B$,
the value of the spin gradually become a monotonically decreasing function of the warp radius. Therefore, for a black hole with fixed spin $a$, as the value of the parameter $B$ is in the intermediate region, we find that the same precession period can be shared by two orbits with different warped radii.
This degeneracy of precession periods is not appeared in the usual Kerr black hole case without magnetic parameter $B$. Moreover, the allowed region in the space of the warp radius and black hole's spin decreases with the parameter $B$ in this intermediate region.
\begin{figure}[ht]
\includegraphics[width=15cm]{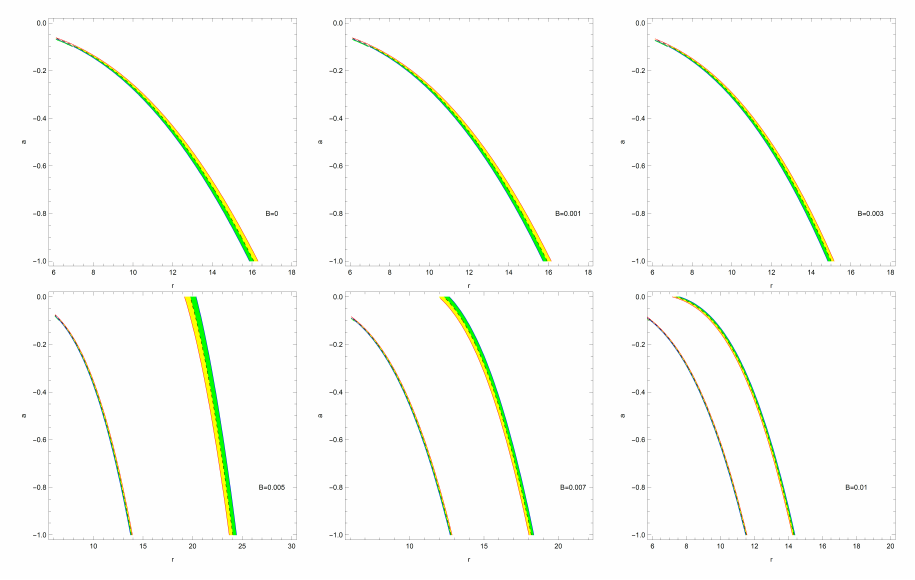}
\caption{Constrains of the black hole spin and warp radius of the accretion disk with the precessing period of the jet nozzle of M87* for different magnetic field parameter $B$ in the retrograde cases. The tilt angle $\xi=\pi/144=1.25^{\circ}$. The solid curve corresponds the period $T=11.24$ years. The blue and red dashed curves are for $T=10.77$ and $11.71$ years.}
\label{sx6}
\end{figure}

In Fig. \ref{sx6}, we analyze the retrograde case. As the parameter $B$ is small, we also find that for each fixed period the absolute value of the black hole spin increases with the warp radius as in \cite{constraining}. However, as $B$ increases up to certain value, we find that there exist two disconnect allowed regions in the space consisting of the warp radius and black hole's spin.
With the increase of $B$, these two regions gradually become closer. In each panel, the precession period increases with the warp radius in the left allowed region and decreases in the right allowed region. In the left allowed region, one can find that the $a>-0.07$ is excluded, which means that the allowed range of $a$ is less than that in the right allowed region.
 With the increase of the parameter $B$, we observed that the constraints imposed by the observation becomes tighter in the left region, while it becomes looser in the right region. Moreover, it is found that the phenomena of sharing same precession period by two different spherical orbits also exists in the retrogrades case with the large $B$, which is different that in the prograde cases.

\begin{figure}[ht]
\includegraphics[width=6cm]{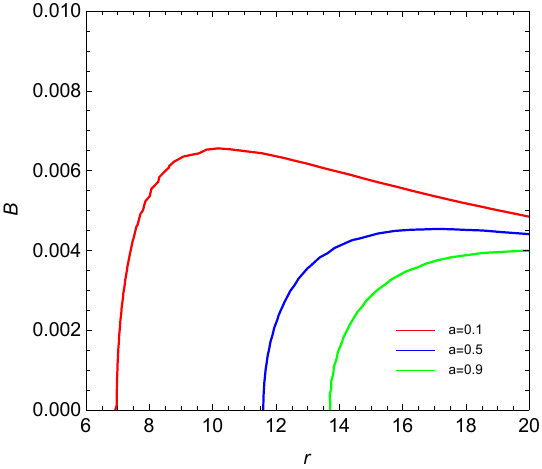}\quad\includegraphics[width=6cm]{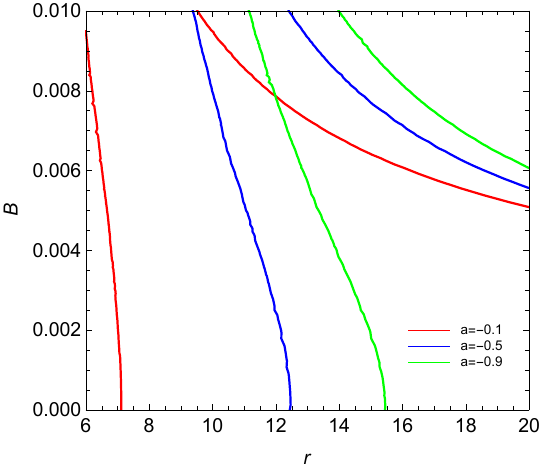}
\caption{Constrains of the magnetic parameter $B$ and warp radius of the accretion disk with the precessing period of the jet nozzle of M87* for different spin parameter $a$ in the retrograde cases. The tilt angle is set to $\xi=\pi/144=1.25^{\circ}$ and the precessing period is fixed to $T=11.24$ years.}
\label{sx8}
\end{figure}

In Fig. \ref{sx8}, we also present changes of the magnetic parameter $B$ with the warp radius for different spin parameter $a$ under the fixed precessing period $T=11.24$ years. For the prograde cases with $a>0$, we find that there is an upper bound for the parameter $B$ and the upper bound decreases with the spin parameter $a$. For the retrograde case, the constraint on the the  magnetic parameter $B$ is looser. Fig. \ref{sx8} also confirms that the range of existing
degeneracy of precession periods in the parameter space $a-B$ is more broaden in the retrograde case, which is consistent with the previous analyses.

Theoretically, effects of the magnetic field on the precession periods of jets could yield certain potential observable signals for astrophysical black holes.
However, for the black hole $M87^*$, its the magnetic field strengthen is estimated to be less than 30 $Gs$, i.e., $B\leq 30 Gs$
from the polarization image \cite{EHT1,EHT2} and then the value of the parameter $B$ in natural unit is $B\leq \frac{30}{2.4\times 10^{19}}\frac{M}{M_{\odot}}\sim 0.8\times 10^{-8}$ \cite{superradiant1,superradiant2}, which means that it is difficult to observe the effects from such tiny magnetic field on precession period around the black hole $M87^*$. The astrophysical systems with strong magnetic field could be appropriate candidates to observe such effects arising from the magnetic field. For instance, the Seyfert galaxy MCG-6-30-15 with the magnetic field $B\approx 10^{10}-10^{11} Gs$ and the mass $M=10^8 M_{\odot}$ \cite{Zakharov}, which means that the value of the parameter $B$ in natural unit is $B\approx 0.04-0.4$. Anyway, our study could help to understand properties of rotating black holes immersed in Melvin magnetic field and the relationship between the magnetic field and the precession of jets.

\section{Discussions and conclusions}
\label{Conclusion}

We investigate the spherical orbits of timelike particles moving in the background of a rotating black hole immersed in the Melvin magnetic field, and probe effects of the magnetic field on the precession period. Our results show that the presence of the magnetic field yields that both the orbital energy $E$ and the angular momentum's absolute values $|L|$ of particles increase in the prograde or retrograde cases. And the effects of magnetic field on $|L|$ of particles moving along the prograde orbits are larger than those along the retrograde orbits. With the increase of the tilt angle $\xi$ of orbital planes,  these effects of the magnetic field gradually decrease on and finally vanish as $\xi=\frac{\pi}{2}$. Moreover, the precession becomes rapidly with the increase of the  magnetic field parameter. Modelling the spherical orbit to the warp radius in the accretion disk and with  the observed precession period from M87*, we present the allowed regions of the black hole spin $a$ and the warp radius $r$ in the accretion disk for different magnetic field parameter $B$.

As the parameter $B$ is smaller, one can find that for each fixed period the absolute value of the black hole spin increases with the warp radius in both the prograde and retrograde cases, which is similar to that in the Kerr case. For the prograde cases, with the gradual increasing of the parameter $B$, the value of the black hole spin first increases and then decreases with the warp radius, and finally become a monotonically decreasing function of the warp radius. For the retrograde  cases, with the gradual increasing of the parameter $B$, we find that there exist two disconnect allowed regions in the space consisting of the warp radius and black hole's spin and these two regions gradually become closer for the larger $B$. And the constraints imposed by the observation becomes tighter in the left region, while it becomes looser in the right region. Especially, we find a novel degenerated phenomenon of precession periods for two different spherical orbits, which does not appear in the usual Kerr black hole case.
The degenerated phenomena of precession periods appear in the prograde case with as the value of the parameter $B$ is in the intermediate region, but in the retrogrades case,  such kind of degenerated phenomena also appear in the case with the large $B$. We also discuss the possibility of observing
effects of the magnetic field on the precession periods of jets for astrophysical black holes. Our study could help to understand properties of rotating black holes immersed in Melvin magnetic field and the relationship between the magnetic field and the precession of jets.

\section*{Acknowledgements}
This work was supported by the National Key Research and Development Program of China (Grant No. 2020YFC2201400) and National Natural Science Foundation of China (Grant Nos. 12275079 and 12035005), and the innovative research group of Hunan Province under Grant No. 2024JJ1006.


\begin{thebibliography}{99}

\bibitem{Akiyama}
 K. Akiyama et al. [Event Horizon Telescope Collaboration],
 {\em First M87 Event Horizon Telescope Results. I. The Shadow of the Supermassive Black Hole},
  Astrophys. J. \textbf{875}, L1 (2019), [arXiv:1906.11238 [astro-ph.GA]].
  
\bibitem{Akiyamab}
 K. Akiyama et al. [Event Horizon Telescope],
 {\em First Sagittarius A* Event Horizon Telescope Results. I. The Shadow of the Supermassive Black Hole in the Center of the Milky Way},
 Astrophys. J. Lett. \textbf{930}, L12 (2022), [arXiv:2311.08680 [astro-ph.HE]].

\bibitem{Kocherlakota}
 P. Kocherlakota et al. [Event Horizon Telescope],
 {\em Constraints on black-hole charges with the 2017 EHT observations of M87*},
 Phys. Rev. D \textbf{103}, 104047 (2021), [arXiv:2311.08680 [astro-ph.HE]].

\bibitem{kerr1} J. M. Bardeen, \textit{in Black Holes (Les Astres Occlus)}, edited by C. DeWitt and B. DeWitt (Gordon and Breach, New York, 1973), p. 215-239.
\bibitem{kerr2} S. Chandrasekhar, \textit{The Mathematical Theory of Black Holes} (Oxford University Press, New York, 1992).

\bibitem{bhs0}Pedro V. P. Cunha, Carlos A. R. Herdeiro, \textit{ Shadows and strong gravitational lensing: a brief review}, Gen. Rel. Grav. {\bf50},  42 (2018).
\bibitem{bhs1} V. Perlick, O. Yu, Tsupko\textit{Calculating black hole shadows: Review of analytical studies}, Phys. Rep. {\bf947}, 1 (2022).
\bibitem{bhs2} S. Chen, J. Jing, W. Qian, and B. Wang, \textit{Black hole images: A review}, Sci. China Phys. Mech.
Astron. {\bf66}, 260401 (2023).
\bibitem{bhs3} M. Wang, S. Chen, J. Jing, \textit{Chaotic shadows of black holes: a short review}, Commun. Theor. Phys. {\bf74}, 097401 (2022).

\bibitem{bhsp1} X. Liu, S. Chen, and J. Jing, \textit{Polarization distribution in the image of a synchrotron emitting ring
around a regular black hole}, Sci. China Phys. Mech. Astron. {\bf65}, 120411(2022).



\bibitem{sw} P. V. P. Cunha, C. A. Herdeiro, E. Radu and H. F. Runarsson, \textit{Shadows of Kerr black holes with scalar hair }, Phys. Rev. Lett. {\bf115}, 211102 (2015), [arXiv:1509.00021];
\bibitem{swo} P. V. P. Cunha, C. A. Herdeiro, E. Radu and H. F. Runarsson, \textit{Shadows of Kerr black holes with and without scalar hair}, Int. J. Mod. Phys. {\bf25}, 1641021 (2016), [arXiv:1605.08293].
\bibitem{astro}F. H. Vincent, E. Gourgoulhon, C. Herdeiro and E. Radu, \textit{Astrophysical imaging of Kerr black holes with scalar hair}, Phys. Rev. D {\bf94}, 084045 (2016), [arXiv:1606.04246].
\bibitem{chaotic} P. V. P. Cunha, J. Grover, C. Herdeiro, E. Radu, H. Runarsson, and A. Wittig, \textit{Chaotic lensing around boson stars and Kerr black holes with scalar hair}, Phys. Rev. D {\bf94}, 104023 (2016).
\bibitem{Banerjee}
 I. Banerjee, S. Chakraborty, and S. SenGupta,
 {\em Silhouette of M87*: A new window to peek into the world of hidden dimensions},
     Phys. Rev. D \textbf{101}, 041301 (2020),
 [arXiv:1909.09385 [gr-qc]].



\bibitem{Lu}
 R.-S. Lu et al.,
 {\em A ring-like accretion structure in M87 connecting its black hole and jet},
     Nature \textbf{616}, 686 (2023), [arXiv:2304.13252 [astro-ph.HE]].

\bibitem{Yuzhu}
 Y. Cui et al,
 {\em Precessing jet nozzle connecting to a spinning black hole in M87},
 Nature \textbf{621}, 711 (2023), [arXiv:2310.09015 [astro-ph.HE]].

\bibitem{Lodato}
 G. Lodato and D. Price
 {\em On the diffusive propagation of warps in thin accretion discs},
     Mon. Not. R. Astron. Soc. \textbf{405}, 1212 (2010), [arXiv:1002.2973 [astro-ph.HE]].


\bibitem{Petterson}
 J. A. Petterson,
 {\em Twisted accretion disks. I. Derivation of the basic equations},
  Astrophys. J. \textbf{214}, 550 (1977).


\bibitem{Ostriker}
 E. C. Ostriker and J. J. Binney,
 {\em Warped and tilted galactic discs},
         Mon. Not. Roy. Astron. Soc. \textbf{237}, 785 (1989).
 \bibitem{Zahrani}  A. M. A. Zahrani,
 {\em Tilted Circular Orbits around a Kerr Black Hole},
     Phys. Rev. D \textbf{109}, 024029 (2024),
 [arXiv:2312.12988 [gr-qc]].
        
 
\bibitem{Fragile}
 P. C. Fragile, G. J. Mathews, and J. R. Wilson,
 {\em Bardeen-Petterson effect and quasi-periodic oscillations in $x$-ray binaries},
     Astrophys. J. \textbf{553}, 955 (2001), [arXiv:astro-ph/0007478 [astro-ph]].




\bibitem{constraining}
 Shao-W. W, Yuan-C. Z, Yu.P.Z, Yu.X.L, \textit{Constraining black hole parameters with the precessing jet nozzle of M87* }[arXiv:2401.17689 [gr-qc]].
EHT1,EHT2

\bibitem{EHT1} K. Akiyama, et al. (The Event Horizon Telescope Collaboration), \textit{First M87 Event Horizon Telescope Results. VII. Polarization of the Ring}, Astrophys. J. Lett. {\bf910}, L12 (2021), arXiv: 2105.01169.
\bibitem{EHT2} K. Akiyama, et al. (The Event Horizon Telescope Collaboration), \textit{First M87 Event Horizon Telescope Results. VIII. Magnetic Field Structure near The Event Horizon}, Astrophys. J. Lett. {\bf910}, L13 (2021), arXiv: 2105.01173
    
 \bibitem{Zakharov}A. F. Zakharov, N. S. Kardashev, V. N. Lukash and S. V. Repin, \textit{Magnetic fields in AGNs and microquasars}, Mon. Not. Roy. Astron.
Soc. {\bf342}, 1325 (2003) [arXiv:astro-ph/0212008]. 

\bibitem{lv00}  F. J. Ernst, \textit{Black holes in a magnetic universe}, J. Math. Phys. {\bf17}, 54 (1976).
\bibitem{lv01} F. J. Ernst, W. J. Wild, \textit{Kerr black holes in a magnetic universe}, J. Math. Phys. {\bf17}, 182 (1976).


\bibitem{chaom1} V. Karas, D. Vokroulflic\'k, \textit{Chaotic Motion of Test Particles in the Ernst Space-time}, Gen. Rel. Grav. {\bf24}, 7 (1992).
\bibitem{chaom2} M. Wang, S. Chen, J. Jing, \textit{Chaos in the motion of a test scalar particle coupling to the
Einstein tensor in Schwarzschild-Melvin black hole spacetime}, Eur. Phys. J. C {\bf 77}, 208 (2017).


\bibitem{chaomsha1} H. C. D. Lima Junior, P. V. P. Cunha, C. A. R. Herdeiro, and L. C. B. Crispino, \textit{Shadows and lensing of black holes immersed in strong magnetic fields}, Phys. Rev. D {\bf 104}, 044018 (2021). arXiv:2104.09577v1 (2021).
\bibitem{chaomsha2} M. Wang, S. Chen, J. Jing, \textit{Kerr Black hole shadows in Melvin magnetic field with stable photon orbits}, Phys. Rev. D {\bf 104}, 084021  (2021).	arXiv::2104.12304.
  
\bibitem{motta1} S. E. Motta, T. M. Belloni, L. Stella, T. Muoz-Darias, R. Fender, \textit{Precise mass and spin measurements for a stellar-mass black hole
through X-ray timing: the case of GRO J1655-40}, Mon. Not. R. Astron. Soc. {\bf437}, 2554 (2014). arXiv:1309.3652 [astro-ph.HE]
\bibitem{motta2} S. E. Motta, T. Muoz-Darias, A. Sanna, R. Fender, T. Belloni, L. Stella, \textit{Black hole spin measurements through the relativistic precession model: XTE J1550-564}, Mon. Not. R. Astron. Soc. {\bf439}, 65 (2014). arXiv:1312.3114 [astro-ph.HE]
\bibitem{motta3} P. Casella, T. Belloni, L. Stella, \textit{The ABC of low-frequency quasiperiodic oscillations in black-hole candidates: analogies with Zsources}, Astrophys. J. {\bf629}, 403 (2005).
    
\bibitem{motta4} L. Stella, M. Vietri, \textit{Lense-Thirring precession and QPOs in low mass X-ray binaries}, Astrophys. J. {\bf 492}, L59 (1998); arXiv: astro-ph/9709085.

  
\bibitem{superradiant1} R. A. Konoplya,  \textit{Magnetic field creates strong superradiant instability}, Phys. Lett. B {\bf666},  283 (2008).
\bibitem{superradiant2} R. Brito, V. Cardoso, P. Pani, \textit{Superradiant instability of black holes immersed in a magnetic field}, Phys. Rev. D {\bf89}, 104045 (2014).






\end{thebibliography}
\end{document}